\documentclass{svproc}
\usepackage{url}
\usepackage{amsmath}
\usepackage[sort&compress]{natbib}
\usepackage{braket}
\usepackage{graphicx}
\usepackage{amssymb}

\begin{document}
\mainmatter              
\title{The lattice Schwinger model and its quantum simulation}
\titlerunning{Lattice Schwinger model}  
%
\author{Jo\~{a}o C. Pinto Barros\inst{1} \and Pierpaolo Fontana\inst{2}
\and Pasquale Sodano\inst{3} \and A. Trombettoni\inst{4}}
\authorrunning{Jo\~{a}o C. Pinto Barros et al.} 
%
\tocauthor{Jo\~{a}o C. Pinto Barros, Pierpaolo Fontana, Pasquale Sodano, and A. Trombettoni}
\institute{Institut f\"ur Theoretische Physik, ETH Z\"urich, Wolfgang-Pauli-Str. 27, \\
8093 Z\"urich, Switzerland\\
\and
Departament de Física, Universitat Aut\`{o}noma de Barcelona, \\
08193 Bellaterra, Spain
\and
INFN, Sezione di Perugia, Via A. Pascoli, 
06123
Perugia, Italy
\email{pasquale.sodano01@gmail.com}
\and
Department of Physics, University of Trieste, 
and INFN, Sezione di Trieste, 
Strada Costiera 11, 
34136 Trieste, Italy
}

\maketitle            

\begin{abstract}
In this chapter we review results on the lattice Schwinger model. In particular, we show how the effect of the anomaly are reproduced on the lattice. We connect these remarkable results to recent developments in the field of quantum simulation of interacting field theories. Schemes for the quantum simulation of (approximations of) Schwinger models are discussed.  
\end{abstract}

\section{Introduction}

It is a pleasure to contribute to a volume celebrating the many and relevant contributions given by Professor Gordon W. Semenoff to modern theoretical physics. 
One of the authors (PS) had the chance to meet Gordon when he and Gordon were very young graduate students in the theoretical physics department of the University of Alberta. We started to discuss physics very soon helped in that by the fact that we shared the same Ph.D. supervisor, Prof. Hiroomi Umezawa. Since then most of my career benefited from the stimulating collaboration with Gordon and my life in general enjoyed our friendship. Many anecdotes come to my mind, but I hope that there will be a chance to share them around a dinner table.

The contribution to this volume is centered on the lattice Schwinger model with staggered fermions. The Schwinger model is the one-flavor quantum electrodynamics (QED) in one spatial dimension. Historically, Schwinger considered this model \cite{Schwinger62bis} as an example of a gauge vector field that can have a non-zero mass \cite{Schwinger1962}. Despite its simplicity as an Abelian gauge theory in one spatial dimension, it has many interesting physical features like, for instance, 
anomalies \cite{ANO}, confinement \cite{Frishman2009} and chiral symmetry breaking \cite{b60}. This made the model very attractive to test analytical and numerical methods relevant for quantum field theory.
Recently, analyses of the Schwinger model were motivated by a growing interest in its connections with quantum simulators \cite{wiese2013}, machine learning \cite{lin2024realtime}, matrix product states \cite{Angelides23} and quantum computing \cite{Funcke}.

The Schwinger model is an ideal laboratory to study the occurrence of anomalies in a field theory. In general, anomalies occur when all the symmetries of a classical field theory can not be realized simultaneously in the corresponding quantum field theory. The classic example is the Adler-Bell-Jackiw anomaly in a vector-like gauge theory, where gauge invariance and global axial symmetries are incompatible. In computations, this phenomenon can be seen as the impossibility of finding a regularization of the ultraviolet divergences which preserves both axial and gauge invariance. If the latter is preserved, the observable manifestation of the anomaly is the failure of conservation laws for axial currents and the resulting absence of their consequences in the spectrum of the quantized theory. The Schwinger model is the prototypical example of a solvable model where all of this occurs.

In the following we shall mainly focus on the one- and two-flavor Schwinger models, reviewing their lattice formulations using staggered fermions and the Hamiltonian approach to lattice gauge theories \cite{b56}. Even though the continuous axial symmetry is broken explicitly by the staggered fermions, the discrete axial symmetry remains and appears in the lattice theory as a translation by one site. An interesting issue which arises in the lattice regularization of gauge theories is how the lattice theory produces the effects of the axial anomaly. 

Lattice field theories are manifestly gauge invariant by construction. Therefore, to properly recover the corresponding continuum theories, they must include a mechanism to violate axial current conservation. Typically, in lattice gauge theories axial anomalies are either cancelled by fermion doubling or the lattice regularization breaks the axial symmetry explicitly. However, in the lattice Schwinger model formulated in the Hamiltonian formalism with one species of staggered fermions, neither of these scenarios occurs. 
Indeed, the effects of the anomaly are not cancelled by doubling since the continuum limit of $(1+1)$-dimensional staggered fermions produces exactly a Dirac fermion, which is the matter content of the model. Furthermore, even if the continuum axial symmetry is explicitly broken, the remaining discrete axial symmetry must be spontaneously broken. The mass operator $\bar\psi\psi$ is odd under discrete axial transformations; if this symmetry is unbroken, then$\braket{\bar\psi\psi}=0$. However, in the continuum theory it is known that \cite{b57_1} $\braket{\bar\psi(x)\psi(x)}=-e^\gamma e_c/2\pi \sqrt{\pi}$, where $\gamma=0.577...$ is the Euler constant and $e_c$ is the electric charge.
This expectation value is called the chiral condensate. The chiral condensate is analogous to the magnetization in a spin model, with the mass $m$ playing the role of the magnetic field. Therefore, even when studying a massless model, determining the sign of the chiral condensate requires the presence of a small mass term $m$ which is subsequently sent to zero. The minus sign appearing in $\braket{\bar\psi(x)\psi(x)}$ arises when taking the limit $m\longrightarrow 0^{+}$ in the Schwinger model Lagrangian. The chiral condensate is non-zero even on the lattice, signaling the spontaneous breaking of discrete axial symmetry. In addition to the analysis of the chiral symmetry breaking pattern on the lattice, we show results rapidly converging to the continuum limit by performing a strong coupling expansion around a gauge invariant ground state \cite{b56}. Recent results on the Kogut-Susskind Hamiltonian approach with staggered fermions \cite{kogut1975hamiltonian} include the study of the discrete chiral symmetry and mass shift \cite{Dempsey2022PRR} and of the real time dynamics and confinement in presence of an Abelian $\mathbb{Z}_n$ gauge group \cite{Magnifico2020realtimedynamics}. 
The two-flavor Schwinger model has been also thouroughfully investigated \cite{Coleman1976,Georgi2020}, with new findings for the phase diagram presented in \cite{Dempsey2023}.

The one-flavor Schwinger model has also been studied as an example of how quantum link models \cite{horn1981,wiese2013} can reproduce the physics of Hamiltonian lattice gauge theories. This versatility made the Schwinger model and its variations as a benchmark for the developments of schemes of quantum simulations of field theories \cite{wiese2013}. The general goal is to physically implement, e.g., in ultracold atoms setups, an interacting system whose low energy description coincides with a target field theory. From one side one can study the effect of the space discretization -- since in typical quantum simulations schemes the used non-relativistic atoms are in optical lattices -- and as well of truncation of Hilbert space,  as it is the case when one implements a discretization of a continuous symmetry. From the other side, one can use the remarkable amount of results from interacting field theories to test and improve the quantum simulation scheme at hand. It is clear that the Schwinger model is extremely useful in both directions and its properties have been widely revisited in the realm of quantum simulations.

We comment on these recent developments in Section \ref{sec:QS}, while Sections \ref{sec:1F} and \ref{sec:2F} are devoted to reviews of relevant features of the Schwinger model on the lattice with one and two flavors. Our presentation partially follow the lines of the discussion presented in the two PhD Theses 
\cite{tesi_Berruto} and \cite{jpbthesis}, the former by Federico Berruto, who worked during his PhD project with Gordon W. Semenoff within the collaboration between University of Perugia and University of British Columbia. 

\section{The one-flavor Schwinger model}
\label{sec:1F}
The continuum Schwinger model \cite{Schwinger62bis,Schwinger1962} stands as the prototypical example of a solvable model where the anomaly occurs. This model is defined as $1+1$-dimensional QED with a single charged, massless Dirac spinor field. The action is 
\begin{equation}
    S = \int d^{2} x \left(\bar{\psi}\left(i\gamma_{\mu}\partial^{\mu}+\gamma_{\mu}A^{\mu}\right)\psi-\frac{1}{4e^{2}_{c}}F_{\mu\nu}F^{\mu\nu}\right)\, .
\end{equation}
It is invariant under the gauge transformations $\psi(x)\to e^{i\theta(x)}\psi(x)$, $\psi^{\dag}(x)\to \psi^{\dag}(x)e^{-i\theta(x)}$, $A_{\mu}\to A_{\mu}+\partial_{\mu}\theta(x)$, and, formally, under the axial phase rotation $\psi(x)\to e^{i\gamma_{5} \alpha} \psi(x)$,
$\psi^{\dag}(x)\to \psi^{\dag}(x) e^{-i\gamma_{5} \alpha}$, where the fifth gamma matrix $\gamma^{5}=i\gamma^{0}\gamma^{1}$ has been introduced. 

At the classical level the above symmetries lead to conservation laws for the vector and axial currents, defined respectively as $j^{\mu}(x)=\bar{\psi}(x)\gamma^{\mu}\psi(x)$ and $j_{5}^{\mu}(x)=\bar{\psi}(x)\gamma^{5}\gamma^{\mu}\psi(x)$. At the quantum level, both currents cannot be simultaneously conserved. If the chosen regularization is gauge invariant, ensuring the conservation of the vector current, then the axial current obeys the anomaly Eq. \cite{b60} 
\begin{equation}
    \partial_{\mu} j_{5}^{\mu}(x)=\frac{e^{2}_{c}}{2\pi}\epsilon^{\mu \nu}F_{\mu\nu}(x)\, ,
\end{equation}
where $\epsilon^{\mu\nu}$ is the two-dimensional Levi-Civita tensor. Furthermore, the correlation functions of the model do not exhibit axial symmetry.

In the spectrum of the Schwinger model, only neutral particles exist. Notably, the bosonic bound state created by the axial current operator, i.e., $\partial^{\mu}\phi(x)=\sqrt{\pi}j_{5}^{\mu}(x)$, appears as a free pseudo-scalar with mass $m_\phi=e_{c}/\sqrt{\pi}$.

Using the bosonic field $\phi$, the action reads as
\begin{equation}
    S=\int d^{2}x\left(\frac{1}{2}\partial_{\mu}\phi
    \partial^{\mu}\phi+\frac{e^{2}_{c}}{2\pi}\phi^{2}\right)\, ,
\end{equation}
and the correlation functions for the vector and axial currents are obtained from the correlation functions of derivatives of the corresponding bosonic fields. The coupling constant $e_{c}$ has the dimension of a mass, and the model is super-renormalizable. 

The Schwinger model has often been used as a field theory where methods for lattice gauge theories, including the strong coupling expansion, can be tested and compared with exact results of the continuum model \cite{b58_1}. 

\subsection{Lattice Schwinger model}
We now revisit the lattice quantization of the Schwinger 
model to show how the effects of the anomaly manifest through spontaneous symmetry breaking.  A lattice Hamiltonian and constraint reproducing the corresponding ones in the continuum limit are:  
\begin{equation}
    H_{S}=\frac{e^{2}_{L}a}{2}\sum_{x}E_{x}^{2}-\frac{it}{2a}\sum_{x}(\psi_{x+1}^{\dag}e^{iA_{x}}\psi_{x}-\psi_{x}^{\dag}e^{-iA_{x}}\psi_{x+1})\, ,
    \label{hamilton1}
\end{equation}
\begin{equation}
    E_{x}-E_{x-1}+\psi_{x}^{\dag}\psi_{x}-\frac{1}{2}= 0\, .
    \label{gauss1}
\end{equation}
Here, the fermionic fields are defined on the sites $x=-\frac{N}{2},-\frac{N}{2}+1,...,\frac{N}{2}$ of a one-dimensional lattice of length $N$ and lattice spacing $a$. The gauge and the electric fields, $ A_{x}$ and $E_{x}$, respectively, are defined on the links joining two adjacent sites $x$ and $x + 1$. The length $N$ is an even integer and, when $N$ is finite, periodic boundary conditions are considered. In the latter case, the continuum limit corresponds to the Schwinger model on a circle \cite{b57_1}.

The coefficient $t$ in the hopping term of Eq. \eqref{hamilton1} plays the role of the speed of light on the lattice. In the naive continuum limit, $e_L=e_c$ and $t=1$. However, it is convenient to leave $e_L$ and $t$ as parameters that can be adjusted to match the lattice values of quantities such as the mass gap and the chiral condensate with those known in the continuum.

Both the Hamiltonian and gauge constraint exhibit the following discrete symmetries: parity $P$, discrete axial symmetry $\Gamma$ and charge conjugation $C$, see, e.g., \cite{b56}. When $A_x=0$, the spectrum of the hopping Hamiltonian in momentum space is $\epsilon(pa)= t \sin pa$, where $p\in 
[0,2 \pi/a]$. In the low-energy limit, this represents a massless relativistic spectrum for excitations near the Fermi level. 
The two intersections of the energy band with the Fermi level provide the continuum right and left moving massless fermions. The electron ground state is invariant under charge conjugation only when the Fermi level is at $\epsilon(p_F)=0$, i.e., when exactly half of the fermion states are filled and $\sum_x\braket{\rho(x)}=0$.  In the remainder of this Section, we focus solely on this scenario, known as the half-filling case.

The lattice Schwinger model is equivalent to a one-dimensional quantum Coulomb gas on the lattice. To discuss this equivalence, it is convenient to employ the Coulomb gauge $A_{x} = A$. By eliminating the non-constant electric field and using the gauge constraint, the following effective Hamiltonian
\begin{eqnarray}
    H_{S}=H_u+H_p
    &\equiv&\left[\frac{e^{2}_{L}}{2 N}E^{2}+\frac{e^{2}_{L}a}{2}
    \sum_{x,y}\rho(x) V(x-y)\rho(y)\right]\nonumber\\
    &+&\left[
    -\frac{it}{2a}\sum_{x}(\psi_{x+1}^{\dag}e^{iA}\psi_{x}-\psi_{x}^{\dag}e^{-iA}
    \psi_{x+1})\right]
    \label{hs1}
\end{eqnarray}
is obtained, where the charge density is defined as $\rho(x)\equiv\psi^{\dag}_x\psi_x-\frac{1}{2}$, and the potential
\begin{equation}
V(x-y)\equiv\frac{1}{N}
\sum^{N-1}_{n=1} e^{i 2\pi n (x-y)/N}\frac{1}{4\sin^2\frac{\pi n}{N}}
\label{cpo}
\end{equation}
represents the Fourier transform of the inverse Laplacian on the lattice for non-zero momentum. Here, the constant electric field is normalized so that $[ A, E ] = i$. The constant modes of the gauge field decouple in the thermodynamic limit $ N \to \infty $.

The gauge fixed Hamiltonian \eqref{hs1} can also be represented as a quantum spin model using the Jordan-Wigner transformation \cite{Korepin} mapping the operator $\psi$'s to spin $S$'s. Consequently, the Hamiltonian is 
\begin{align}    
    H=&\frac{t}{2a}\sum_x\left[e^{iA}S^+(x+1)S^-(x)+e^{-iA}S^-(x+1)S^+(x)\right]\\
    &+\frac{e_L^2a}{2}\sum_{x,y}S^3(x)V(x-y)S^3(y)+\frac{e_L^2E^2}{2N}\, .
    \label{spinham}
\end{align}
For now, we disregard the constant modes $E$ and $A$. The first term in the Hamiltonian is the quantum $XY$ model, known to have a disordered ground state (the exact solution of this model can be obtained by the Jordan-Wigner transformation). The second term is a long-range Ising interaction \cite{Defenu20,Defenu2023} resulting from the infinite range of the Coulomb interaction. When both terms are present in the Hamiltonian, they compete with each other: the first one favors disorder, while the second one promotes order.  

\section{The two-flavor Schwinger model}
\label{sec:2F}
In this Section we discuss the $SU(2)$-flavor lattice Schwinger model in the Hamiltonian formalism using staggered fermions.  The presence of the continuum internal isospin symmetry makes the model considerably more interesting than the one-flavor case; the spectrum is richer, exhibiting also massless excitations. Furthermore, the chiral symmetry breaking pattern differs from the one-flavor case.  One is able to show \cite{b36,b36b} that the strong coupling limit of the two-flavor lattice Schwinger model is mapped onto the one-dimensional spin-$1/2$ quantum Heisenberg antiferromagnet. The ground state of the antiferromagnetic chain and the complete spectrum have been known since many years \cite{Korepin}.  

The action of the $1+1$-dimensional electrodynamics with two charged 
Dirac spinor fields is
\begin{equation}
S = \int d^{2} x\left[\sum_{a=1}^{2} \overline{\psi}_{a}
(i\gamma_{\mu}\partial^{\mu}+\gamma_{\mu}
A^{\mu})\psi_{a}-\frac{1}{4e^{2}_{c}}F_{\mu\nu}F^{\mu\nu}\right]\, .
\label{action}
\end{equation}
The theory has an internal $SU_L(2)\otimes SU_R(2)$-flavor isospin symmetry ($L$ and $R$ standing for left and right); the Dirac fields are an isodoublet whereas the electromagnetic field is an isosinglet.

A lattice Hamiltonian and constraint are 
\begin{equation}
H_{S}=\frac{e^{2}a}{2}\sum_{x=1}^N E_{x}^{2}-\frac{it}{2a}\sum_{x=1}^N
\sum_{a=1}^2 \left(\psi_{a,x+1}^{\dag}e^{iA_{x}}\psi_{a,x}
-\psi_{a,x}^{\dag}e^{-iA_{x}}\psi_{a,x+1}\right)\, ,\label{hamilton}
\end{equation}
\begin{equation}
E_{x}-E_{x-1}+\psi_{1,x}^{\dag}\psi_{1,x}+\psi_{2,x}^{\dag}\psi_{2,x}-1 
=0\, .
\label{gauss}
\end{equation}
The fermion fields are defined on the sites, $x=1,...,N$, the gauge and electric fields, $A_{ x}$ and $E_{x}$, on the links $[x; x + 1]$, $N$ is an even integer and, when $N$ is finite it is convenient to impose periodic boundary conditions. When $N$ is finite, the continuum limit is the two-flavor Schwinger model on a circle. The coefficient $t$ of the hopping term in (\ref{hamilton}) plays the role of the lattice light speed. In the naive continuum limit, $e_L=e_c$ and $t=1$. 

The Hamiltonian and gauge constraint exhibit the discrete symmetries parity $P$, axial symmetry $\Gamma$, charge conjugation $C$ and as well $G$-parity: $A_{x} \to -A_{x+1}$, $E_{x} \to -E_{x+1}$, $\psi_{1,x} \to \psi^{\dag}_{2,x+1}$, 
$\psi_{1,x}^{\dag}
\to \psi_{2,x+1}$, $\psi_{2,x}\to -\psi^{\dag}_{1,x+1}$ $\psi_{2,x}^{\dag}
\to -\psi_{1,x+1}$.

The lattice two-flavor Schwinger model is equivalent to a one dimensional quantum Coulomb gas on the lattice with two kinds of particles. To see this one can fix the gauge, $A_{x} = A$ (Coulomb gauge). Eliminating the non-constant electric field and using the gauge constraint, one obtains the effective Hamiltonian
\begin{eqnarray}
H_{S}&=&H_u+H_p
\equiv\left[\frac{e^{2}_{L}}{2 N}E^{2}+\frac{e^{2}_{L}a}{2}
\sum_{x,y}\rho(x) V(x-y)\rho(y)\right]
\nonumber\\
&+&\left[
-\frac{it}{2a}\sum_{x}\sum_{a=1}^{2}(\psi_{a,x+1}^{\dag}e^{iA}\psi_{a,x}-\psi_{a,x}^{\dag}e^{-iA}
\psi_{a,x+1})\right]\ ,
\label{hs}
\end{eqnarray}
where the charge density is $\rho(x)=\psi^{\dag}_{1,x}\psi_{1,x}+\psi^{\dag}_{2,x}\psi_{2,x}-1$, and the potential is again given by
\begin{equation}
V(x-y)=\frac{1}{N}
\sum^{N-1}_{n=1} e^{i 2\pi n (x-y)/N}\frac{1}{4\sin^2\frac{\pi n}{N}}\, .
\end{equation}
The constant modes of the gauge field decouple in the thermodynamic limit $ N \longrightarrow \infty $.

\section{Quantum simulation of the Schwinger model}
\label{sec:QS}

A significant amount of work surged in the last decade on the quantum simulation of interacting field theories \cite{wiese2013,BBM2020,halimeh2023}.  Encoding gauge symmetry in quantum simulation experiments is the focal issue in quantum simulation of gauge theories. The Schwinger model, or variations of it, has been one of the prime targets of the community and has been largely used as a case study.. This activity went along with the remarkable progress in the last years associated with several experiments that have realized a variety of modified versions of the model, with different degrees of complexity.

There are several strategies designed to achieve gauge symmetry or to encode an effective description of the model. We refer to \cite{wiese2013,BBM2020,halimeh2023} for useful material and reviews on quantum simulations of the Schwinger model and beyond. Here for the sake of completeness, 
we will remind two of the main approaches as the literature has investigated so far: energy penalty and integration of the gauge fields. We start with a description of the one-flavor case and then we briefly discuss how these can be generalized for the two-flavor case. 

In the case of energy penalty, gauge symmetry emerges in the low-energy sector of the theory. While the Hilbert space dimension at any matter site is $2$, the Hilbert space dimension of a single gauge link is infinite. Quantum link models (QLM) \cite{horn1981,orland1990} give a way of making this Hilbert space finite while preserving gauge invariance. We will start with a brief review of QLM and show how these models can emerge in the low-energy sector of a theory of one fermionic species coupled to two bosonic degrees of freedom, see \cite{halimeh2023, Fontana23, tesi_Fontana} for more references.

In the second case, gauge fields are integrated out. One advantage of this approach is that no type of truncation is necessary. After removing the gauge degrees of freedom only an effective theory of one fermionic (or spin) degree of freedom per site is left out. The main price to pay is that the effective Hamiltonian is no longer local.

In the rest of the Section we shall denote respectively by $c$'s and $b$'s the fermionic and bosonic operators associated to the (non-relativistic) fermions and bosons with which the quantum simulation is performed. When two species of bosons are needed, they will be denoted by $b^{\left(1\right)}$, $b^{\left(2\right)}$. Moreover, for the one-flavor Schwinger model the abbreviation $    G_x=E_{x}-E_{x-1}-c_{x}^{\dagger}c_{x}+\frac{1-\left(-1\right)^{x}}{2}$ will be used. The operators $G_x$ are the generators of gauge transformations.

QLMs \cite{horn1981,orland1990} ensure that gauge invariance is preserved by maintaining the commutation relations and sacrificing unitarity through the identification with spins
\begin{equation}
    U_x\rightarrow S^+_x,\quad U_x^\dagger\rightarrow S^-_x,\quad 
    E_x\rightarrow S^z_x \, .
    \label{QLM_bosonic_spin_variables}
\end{equation}
In this framework, the local Hilbert space becomes finite-dimensional, more precisely it has dimension $(2S+1)$ for a quantum link of spin-$S$. The operators $U_x$, previously unitary operators, no longer commute: $[U_x,U^\dagger_y]=2\delta_{xy}E_x$.

We now discuss the effective low-energy gauge theory using the energy penalty approach. The idea behind this approach consists on building a Hamiltonian which does not prohibit the symmetry violation from occurring, but instead penalizes it with a large energy. More concretely, let suppose one wants to implement a set of symmetries that has a respective set of generators $\left\{ G_{x}\right\}$ commuting with each other $\left[G_{x},G_{y}\right]=0$. Let $H_{0}$ be a Hamiltonian which does not respect these symmetries. Then one constructs the following Hamiltonian:
\begin{equation}
H=H_{0}+\Gamma\underset{x}{\sum}G_{x}^{2}\, ,
\label{eq:Hpenalty}
\end{equation}
where $\Gamma$ is a large energy scale (much larger than the energy scales involved in $H_{0}$). Since $G_{x}$ are Hermitian
$G_{x}^{2}$ have non-negative eigenvalues. One can choose the lowest eigenvalue to be zero by an appropriate definition of $G_{x}$. Then, at low energy $\left(\ll\Gamma\right)$, the states will respect approximately the condition $G_{x}\left|\psi\right\rangle \simeq0$. It is then possible to construct an effective Hamiltonian, valid in low energy, which will respect the symmetries generated by $\left\{ G_{x}\right\}$.

Let $G$ be the projector operator on the subspace of the total Hilbert space obeying $G_{x}\left|\psi\right\rangle =0$ and let $P=1-G$. Then the low-energy Hamiltonian can be written as:
\begin{equation}
H_{eff}=GH_{0}G-\frac{1}{\Gamma}GH_{0}P\frac{1}{\underset{x}{\sum}G_{x}^{2}}PH_{0}G+{\cal O}\left(\Gamma^{-2}\right)\label{eq:Heff_pert_theory}\, ,
\end{equation}
which respect the symmetries. Within this framework, an effective Abelian gauge theory can be constructed. 

We now address how this can be used to simulate the Schwinger model. The fermionic matter fields will be described by fermionic species. For gauge fields one can use two species of bosons to write
\begin{equation}
U_{x}=b_{x}^{\left(2\right)}{}^{\dagger}b_{x}^{\left(1\right)},\ E_{x}=\frac{1}{2}\left(b_{x}^{\left(2\right)}{}^{\dagger}b_{x}^{\left(2\right)}-b_{x}^{\left(1\right)}{}^{\dagger}b_{x}^{\left(1\right)}\right)\, .\label{eq:Schwinger_bosons}
\end{equation}
Each link is loaded with a total of $2S$ bosons where $S$ is a half-integer. Then one has the desired representation for the quantum links in terms of atomic variables. In the case of atomic quantum simulations one can build a one-dimensional optical lattice where fermions are allowed to hop among lattice points and in each link there are a total of $2S$ bosons. The target Hamiltonian with Schwinger bosons can be written as
\begin{align}
H=-t&\underset{x}{\sum}\left(c_{x}^{\dagger}b_{x}^{\left(\bar{\sigma}\right)}{}^{\dagger}b_{x}^{\left(\sigma\right)}c_{x+1}+\mathrm{h.c.}\right)+m\underset{x}{\sum}\left(-1\right)^{x}c_{x}^{\dagger}c_{x}\nonumber \\
+&\frac{g^{2}}{8}\underset{x}{\sum}\left(b_{x}^{\left(2\right)}{}^{\dagger}b_{x}^{\left(2\right)}-b_{x}^{\left(1\right)}{}^{\dagger}b_{x}^{\left(1\right)}\right)^{2}\, ,\label{eq:H_U(1)_SchwingerBosons}
\end{align}
(using the convention $\sigma=1$ and $\bar{\sigma}=2$). 

The last two terms can be, in principle, implemented directly properly tuning the interactions between the bosons and the potential for the fermions. The first term, instead, is a correlated hopping between bosons and fermions, less easy to implement. Furthermore the terms like $b_{x}^{\left(\bar{\sigma}\right)}{}^{\dagger}b_{x}^{\left(\sigma\right)}$ and $c_{x}^{\dagger}c_{x+1}$, which are not gauge invariant, must be suppressed. This is solved by the energy penalty approach. In general the non-gauge invariant Hamiltonian with the ingredients described has then the form:
\begin{equation}
\begin{split}
H_{0}=&-\underset{x}{\sum}\left[t_{F}\left(c_{x}^{\dagger}c_{x+1}+\mathrm{h.c.}\right)-t_{B}\left(b_{x}^{\left(2\right)}{}^{\dagger}b_{x}^{\left(1\right)}+\mathrm{h.c.}\right)\right]\\
&+\underset{x}{\sum}\left(v_{x}^{F}c_{x}^{\dagger}c_{x}+\underset{\sigma}{\sum}v_{x}^{B\sigma}b_{x}^{\left(\sigma\right)}{}^{\dagger}b_{x}^{\left(\sigma\right)}\right)+U\underset{x}{\sum}\left(b_{x}^{\left(2\right)}{}^{\dagger}b_{x}^{\left(2\right)}-b_{x}^{\left(1\right)}{}^{\dagger}b_{x}^{\left(1\right)}\, .\right)^{2}\label{eq:H0_energypenalty}
\end{split}
\end{equation}
Using the generators for the $U\left(1\right)$ to penalize states out of the physical sector, we can write
\begin{equation}
H=H_{0}+\Gamma\underset{x}{\sum}\left(E_{x}-E_{x-1}-c_{x}^{\dagger}c_{x}+\frac{1-\left(-1\right)^{x}}{2}\right)^{2}\, ,
\label{eq:Hpun}
\end{equation}
showing how the Schwinger model may emerge from the non-relativistic setup. It is crucial that one has access to the interactions that are introduced on the last term corresponding to the energy penalty. 

\begin{figure}
\begin{center}
\includegraphics[width=.7\linewidth]{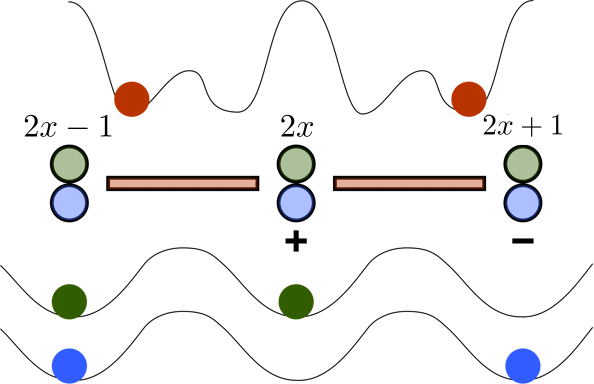}
\end{center}
\par
\caption{Superlattice configurations for the bosonic and fermionic species towards the quantum simulation of the two-flavor Schwinger model. When compared to the one-flavor case, another fermionic species must be added that should also respect the correlated hopping with the bosons on the links.
\label{fig:Superlattice-2flavor}}
\end{figure}

The inclusion of extra flavors can be incorporated in both descriptions above. In the case of energy penalty, this entails introducing a new fermionic species. This implies further fine-tuning concerning fermion-fermion interactions (across species) and fermion-boson interactions with the new species. For the case of integration of gauge fields, one will have an effective spin ladder with long-range interactions. For the energy penalty scheme, there would be an extra fermionic species, which should be added to \eqref{eq:H0_energypenalty}. The charge density at a point $x$ is given by
\begin{equation}
    \rho_x=c^\dagger_{1x}c_{1x}+c^\dagger_{2x}c_{2x}-1+\left(-1\right)^x\, .
\end{equation}
As a consequence, \eqref{eq:Hpun} should be replaced by
\begin{equation}
H=H_{0}+\Gamma\underset{x}{\sum}\left(E_{x}-E_{x-1}-\rho_x\right)^2.
\label{eq:Hpun2}
\end{equation}
This will give rise to new types of interaction to the Hamiltonian but the principles are analogous. A representation of this case is presented in Fig. \ref{fig:Superlattice-2flavor}.

\section*{Acknowledgements} The authors acknowledge discussions with F. Berruto, M. Burrello, A. Celi, M. Dalmonte, N. Defenu, M. C. Diamatini, D. Giuliano, G. Grignani, E. Langmann, L. Lepori, M. K. Marinkovi\'{c}, G. Mussardo, A. Papa and U.-J. Wiese for discussions and collaborations on the different topics discussed in this chapter. Special thanks go of course to Gordon W. Semenoff for a lot of inspiring discussions on lattice gauge theories and many other topics.

\bibliographystyle{unsrt}
\bibliography{Semenoff_Book}

\end{document}